\documentclass[10.5pt,a4paper,aps,prb,preprintnumbers,amsmath,amssymb,twocolumn,longbibliography,superscriptaddress]{revtex4-1}

\usepackage{amsmath}
\usepackage{amssymb}
\usepackage{mathtools}
\usepackage[utf8]{inputenc}
\usepackage[T1]{fontenc}
\usepackage{graphicx}
\usepackage{braket}
\usepackage{physics}
\usepackage{here}
\usepackage{wrapfig}
\usepackage{color}
\usepackage{array}
\usepackage{comment}

\begin{document}

\title{Competing order in two-band Bose-Hubbard chains with extended-range interactions}

\author{Yuma Watanabe}
\affiliation{ICFO-Institut de Ciencies Fotoniques, The Barcelona Institute of Science and Technology, 08860 Castelldefels (Barcelona), Spain}
\author{Utso Bhattacharya}
\affiliation{ICFO-Institut de Ciencies Fotoniques, The Barcelona Institute of Science and Technology, 08860 Castelldefels (Barcelona), Spain}
\author{Ravindra W.  Chhajlany}
\affiliation{Institute of Spintronics and Quantum Information, Faculty of Physics, Adam Mickiewicz University, 61614 Poznan, Poland}
\author{Javier Arg\"{u}ello-Luengo}
\affiliation{ICFO-Institut de Ciencies Fotoniques, The Barcelona Institute of Science and Technology, 08860 Castelldefels (Barcelona), Spain}
\affiliation{Departament de Física, Universitat Politècnica de Catalunya, Campus Nord B4-B5, 08034 Barcelona, Spain}
\author{Maciej Lewenstein}
\affiliation{ICFO-Institut de Ciencies Fotoniques, The Barcelona Institute of Science and Technology, 08860 Castelldefels (Barcelona), Spain}
\affiliation{ICREA, Pg. Lluis Companys 23, 08010 Barcelona, Spain}
\author{Tobias Gra\ss}
\affiliation{DIPC - Donostia International Physics Center, Paseo Manuel de Lardiz{\'a}bal 4, 20018 San Sebasti{\'a}n, Spain}
\affiliation{IKERBASQUE, Basque Foundation for Science, Plaza Euskadi 5, 48009 Bilbao, Spain}
\affiliation{ICFO-Institut de Ciencies Fotoniques, The Barcelona Institute of Science and Technology, 08860 Castelldefels (Barcelona), Spain}

\begin{abstract}
   Motivated by the recent progress in realizing and controlling extended Bose-Hubbard systems using excitonic or atomic devices, the present Letter theoretically investigates the case of a two-band Bose-Hubbard chain with nearest-neighbor interactions. Specifically, this study concentrates on the scenario where, due to the interactions, one band supports a density wave phase, i.e. a correlated insulating phase with spontaneous breaking of translational symmetry in the lattice, while the other band supports superfluid behavior. Using the density matrix renormalization group method, we show that supersolid order can emerge from such a combination, that is, an elusive quantum state that combines crystalline order with long-range phase coherence. Depending on the filling of the bands and the interband interaction strength, the supersolid phase competes with phase-separation, superfluid order, or Mott insulating density-wave order. As a possible setup to observe supersolidity, we propose the combination of a lower band supporting density-wave order and a thermally excited band that supports superfluidity due to weaker lattice confinement.
\end{abstract}
\maketitle

\textit{Introduction.} Bosons have long occupied
a niche role in the study of strongly correlated many-body phases. While interacting fermions in the lattice have been described by the Hubbard model since 1963 \cite{Hubbard1963}, the bosonic variant of this model gained prominence only after 1989 \cite{Fisher1989} (although it was essentially introduced already by Gersch and Knollman in 1963 ~\cite{Gersch1963} to study the effects of repulsive interactions on condensation of bosons). As for fermions, also the Bose-Hubbard model features a quantum phase transition into a Mott insulating phase, caused by the repulsive local interaction. On the other hand, the weakly interacting Bose system condenses into a superfluid. Apart from the intrinsic theoretical interest in these phases, much of the original motivation for the Bose-Hubbard model was to describe the behavior of $^4$He. However, the model found a second more direct and far-reaching 
application as an accurate description of certain ultracold bosonic atoms moving in optical periodic lattices~\cite{Jaksch1998} that was experimentally confirmed spectacularly in 2002~\cite{Greiner2002}. 
The realization of this theoretical model in a synthetic quantum system gave birth to the field of practical quantum simulations and has been followed by a tremendous amount of experimental and theoretical work \cite{Maciej_book, Gross2017}, for instance exploring different lattice geometries, quasiperiodic potentials, synthetic gauge degrees of freedom, and many more. These developments have put an end to bosons' niche existence in the field of quantum many-body physics. 

The standard Bose-Hubbard model is limited to a single energy band, a simplification that tends to be justified if the lattice potential is the dominant energy scale. However, the possibility of activating additional bands significantly enriches the model \cite{Larson2009,Dutta_2015,Xu2016,Li2016,Chen2021}, and has for instance led to the observation of orbital or staggered superfluidity~\cite{Wirth2011,Wang2021,Song2022}. Another extension of the Bose-Hubbard model that has attracted a lot of attention is long-range interactions \cite{Goral2002,Damski2003,Sengupta2005, Batrouni2006, DallaTorre2006, Menotti2007, Trefzger2008, Mishra2009, Trefzger2009, Hauke_2010, Capogrosso-Sansone2010, Trefzger_2011, Dalmonte2011, Sowiifmmode2012, Maik_2012, Rossini_2012, Ohgeo2012, Batrouni2013, Batrouni_2014, Dutta_2015, Pu2023}. The inclusion of nearest-neighbor interactions can lead to density-modulated phases, that is, a spontaneous breaking of the (discrete) translational symmetry of the lattice. Depending on the filling of the lattice, the density-modulation may occur in the Mott insulating regime, forming phases also known as (charge) density wave, Mott crystal or staggered Mott insulator, or checkerboard phase (in 2D). Most striking, however, is the appearance of density modulations in the superfluid regime, where the coexistence of superfluid correlations and translational symmetry breaking leads to a so-called supersolid phase. Another phase of great interest that is stabilized by nearest-neighbor interactions is the Haldane insulator. This gapped phase does not break the translational symmetry, but is characterized by a hidden string order: Particle–hole fluctuations appear in alternating order and are separated by strings of equally populated
sites.

The experimental realization of the novel phases induced by the long-range interaction is one of the frontiers in quantum many-body physics. Cold atoms stand out by their detection opportunities, as string order parameters can be measured through single-site microscopy \cite{Hilker2017}, and different atomic systems with long-range interactions have been developed. In some atomic species, such as erbium, a strong magnetic dipole moment leads to long-range interactions and has enabled the detection of a supersolid phase in a continuous trap \cite{Ilzhoefer2021} and, very recently, also of charge density waves in a lattice system \cite{Su2023}. In systems of polar molecules, the electric dipole moment is the source of strong long-range interactions. Furthermore, by Rydberg-dressing it is possible to synthesize long-range interactions between cold atoms \cite{Zeiher2016, Browaeys2020}. Notably, bosonic many-body phases can also be studied in electronic systems by optically exciting bound electron-hole pairs. These composite bosons, also known as excitons, can possess a large electric dipole moment. In such a system, the Bose-Hubbard model can be realized by assembling a lattice potential via nanolithography on a GaAs double quantum well \cite{Lagoin2022-1}. Alternatively, twisted layers of transition metal dichalcogenides can be used to produce a tunable lattice geometry originating from the twist-induced Moiré pattern \cite{Park2023}. Only a few months after the implementation of standard Bose-Hubbard physics in the excitonic device, the strong dipolar interactions between excitons have also enabled, for the first time, the realization of the extended Bose-Hubbard model \cite{Lagoin2022} in a 2D square lattice. Indirect hints for the formation of a checkerboard solid have been measured.  Interestingly, due to the extremely strong interactions also multi-band effects may play an important role in this system. 

Motivated by these experimental breakthroughs in a variety of different platforms, the present paper takes a theoretical perspective on the combined effect of nearest-neighbor interactions and two-band physics in one-dimensional Bose-Hubbard systems.
To investigate the ground state phases we use the density matrix renormalization (DMRG) group method~\cite{SCHOLLWOCK201196, Itensor2022}.
In particular, we focus on scenarios where the two bands possess sufficiently distinct parameters, such that each band on its own would be in a different phase. Specifically, we are interested in the case where a band with relatively strong nearest-neighbor interactions favors a charge density wave, whereas the other band is in the superfluid regime due to enhanced tunneling. We analyze the interplay of two such bands, induced by density-density inter-band interactions, and demonstrate that supersolid behavior may emerge from such a combination. After introducing the model, we first concentrate on the case where both bands are equally filled, and then study the fate of the supersolid phase as a function of band population.

\begin{figure}[h]
    \centering
    \includegraphics[width=1.0\columnwidth]{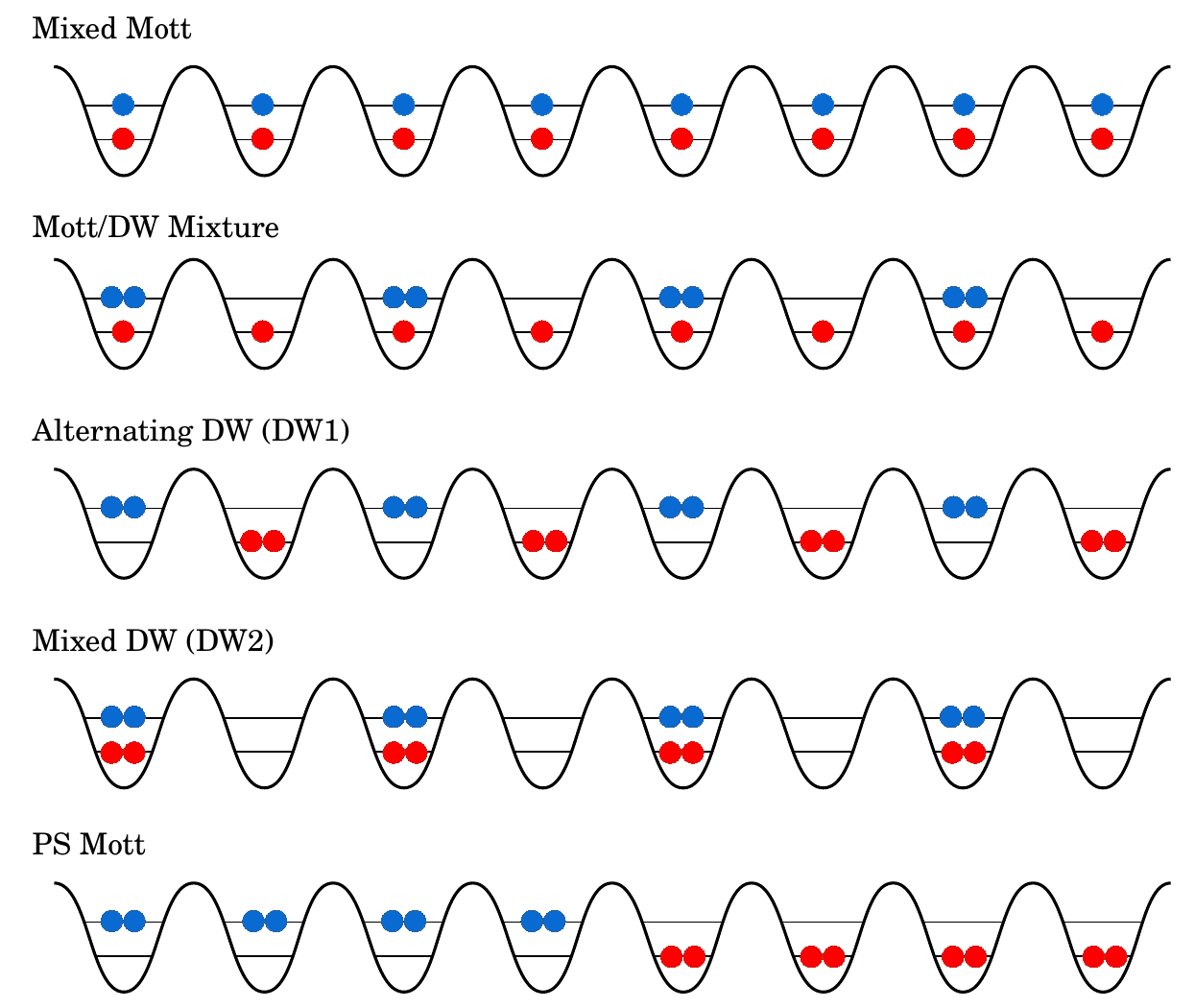}
    \caption{Five possible states at unit-filling with negligibly small hopping. These states appear depending on the competition of interactions. In the case of phase-separation (PS), the two subsystems may form $n=2$ Mott states (as shown here), or, if $V>3U$, density wave patterns with 4-fold occupied sites (not shown).}
    \label{fig:5PossibleState}
\end{figure}

\begin{figure*}[t]
    \centering
    \includegraphics[scale = 0.12]{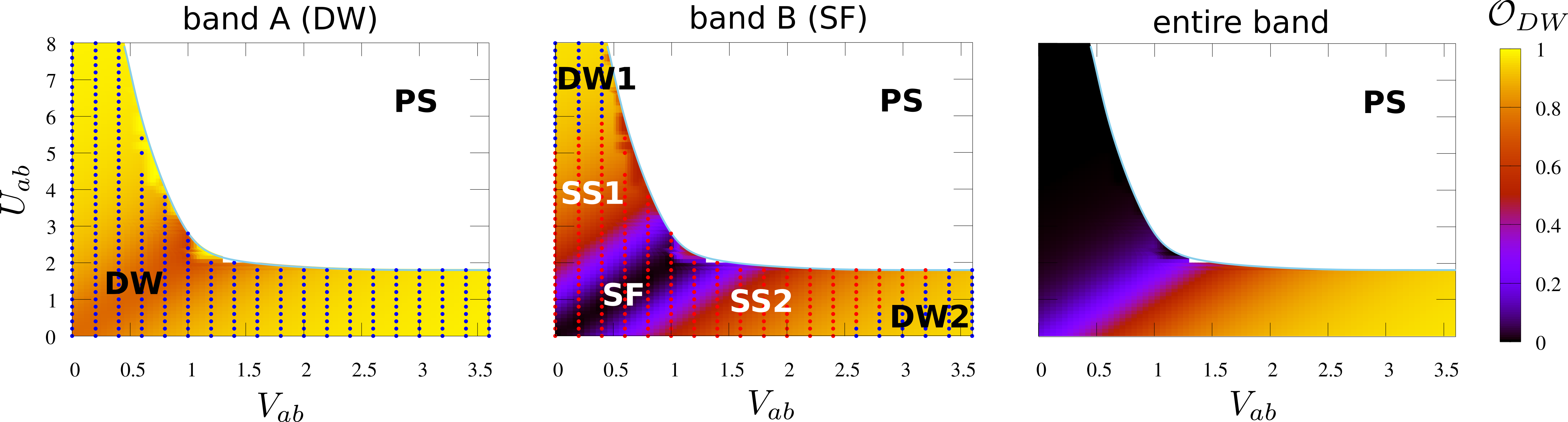}
    \caption
    {
    The ground state phase diagram of an unpolarized system ($L=40$) at unit-filling with $U_a = 5, V_a = 4$ and $U_b = 1, V_b = 0.5$, with $J_a=J_b=1$ being the unit of energy. 
    With this choice, band A is in the DW regime, whereas band B is SF. As a function of interband interactions  $V_{ab}$ and  $U_{ab}$, we identify different phases due to the interplay of the inequivalent bands.
    Specifically, the plots show the DW order parameter (color plot), as well as the decay behavior of the superfluid correlation function (blue or red circles, for exponential or algebraic decay, respectively). 
    We separately consider band A  (left panel) and band B  (middle panel), as well as the entire system jointly (right panel).
    \textcolor{black}{White regions (with the sky-blue line as a guide for the eye), distinguished by inspecting particle number profiles, are identified as phase-separated regimes, whereas the colored regimes exhibit DW and SF, but also supersolid (SS) phases with coexisting SF and DW orders.}
    }
    \label{fig:phasediagram}
\end{figure*}

\textit{Model.} 
We study the one-dimensional two-band extended Bose-Hubbard model described by the following Hamiltonian:
    \begin{align}
    \label{eq:hamiltonian}
        &\hat{H} =
        -J_a \sum_{\Braket{i, j}}( \hat{a}_i^\dagger\hat{a}_j + \texttt{H.c.} )
        -J_b \sum_{\Braket{i, j}}( \hat{b}_i^\dagger\hat{b}_j + \texttt{H.c.} ) \nonumber \\
        & +\frac{U_a}{2}\sum_{i=1}^L \hat{n}_i^a (\hat{n}_i^a - 1)
          +\frac{U_b}{2}\sum_{i=1}^L \hat{n}_i^b (\hat{n}_i^b - 1)
          +U_{ab}\sum_{i=1}^L \hat{n}_i^a\hat{n}_i^b \nonumber \\
        & +V_a \sum_{i=1}^{L-1} \hat{n}_i^a \hat{n}_{i+1}^a
          +V_b \sum_{i=1}^{L-1} \hat{n}_i^b \hat{n}_{i+1}^b \nonumber \\
        & +V_{ab} \sum_{i=1}^{L-1}( \hat{n}_i^a\hat{n}_{i+1}^b + \hat{n}_i^b \hat{n}_{i+1}^a).
    \end{align}
Here, $\hat{a}_i^\dagger$, $\hat{a}_i$ ($\hat{b}_i^\dagger$, $\hat{b}_i$) denote the creation and annihilation operators on site $i$ in band A (B), and $\hat{n}_i^a$  ($\hat{n}_i^b$) denote the corresponding number operators. 
Particles in each band move between neighboring lattice sites with hopping $J_a$ and $J_b$ and interact with each other with strength $U_a$ and $U_b$ in the same sites and $V_a$ and $V_b$ in the neighboring sites. The interplay of the bands is controlled by on-site $U_{ab}$ and nearest neighbor $V_{ab}$ interactions between two bands.
\textcolor{black}{We suppress a possible band gap term which, at a fixed band polarization, reduces to a constant.}

\textit{Unpolarized system.}
Particle number and band polarization are conserved quantities of the Hamiltonian, and we first analyze unpolarized configurations at filling one, e.g.  $N_{a(b)}= \sum_{i=1}^L \hat{n}_i^{a(b)}$ are the populations of the bands, and $L$ is the number of sites.
For the sake of simplicity, we start by considering the small hopping limit $J_a, J_b \rightarrow 0$. In this case, the ground state is determined only by the competition between interactions, and this restriction gives us an intuitive understanding of the individual role of each interaction in our model. In the absence of interband interactions, each band can either form a Mott insulator with one particle per site, if the on-site repulsion $U$ is dominant in this band, or a density wave (DW) phase of alternating doubly occupied and empty sites, if nearest-neighbor interactions are dominant, $V>U/2$.
Five possible combinations of these phases for two-band systems are illustrated in Fig.~\ref{fig:5PossibleState}. Which of these possibilities is chosen is governed by the interband interactions. For instance, if each of the bands favors DW order, the two bands can still be combined to a configuration with uniform total density if the DWs are shifted relative to each other. We note that, nevertheless, translational invariance is broken through the band ordering. If interband nearest-neighbor interactions become strong ($V_{ab}>U_{ab}/2)$, symmetry breaking is seen also in the total density, as a configuration of alternating empty sites and four-fold occupied sites will be chosen. When interband interactions become strong compared to intraband interactions, phase-separated states tend to become favorable, e.g. a splitting of the system in two Mott phases of two identical particles per site.
x

\textcolor{black}{
For a quantitative characterization of DW order, we introduce  normalized correlation functions probing the density order modulated at wave vector $q=\pi$ for each band $\alpha=a,b$ individually, as well as for both bands jointly: 
\begin{gather}
\label{eq:DWcorrelators1}
        C^{\alpha}_{DW}(r) = \frac{(-1)^r}{(\Bar{n}^\alpha)^2} 
       \Braket{\delta\hat{n}_i^\alpha 
       \delta\hat{n}_{i+r}^\alpha}, \\   
\label{eq:DWcorrelators2}
    C_{DW}^{ab}(r) = \frac{(-1)^r}{  
    (\Bar{n}^\alpha+\Bar{n}^b)^2 } 
       \Braket{(\delta\hat{n}_i^a+\delta\hat{n}_i^b) 
        (\delta\hat{n}_{i+r}^a + \delta\hat{n}_{i+r}^b)}, 
\end{gather}
where $\delta\hat{n}_i^\alpha$ is the deviation of the number of particles at site $i$ from  the average particle number in the band $\Bar{n}^\alpha$. 
The DW order parameter is then taken to be the correlation function at the largest distance which in the case of calculations in a finite 
chain of length $L$ is ${\mathcal{O}}_{DW}^\alpha = C_{DW}^{\alpha}(r=L/2)$. 
}
It is known that finite hopping can modify or destroy the insulating states, giving rise to the Haldane insulating phase characterized by topological string order in a one band Bose Hubbard model, or the superfluid phase with long-range phase coherence. In this letter, we are primarily interested in the interplay of density waves and superfluids. 
\textcolor{black}{To characterize the latter, we introduce the superfluid (SF) correlation function defined as
\begin{gather}
        C^{a}_{SF}(r) =\Braket{\hat{a}_i^\dagger \hat{a}_{i+r}}, 
        C^{b}_{SF}(r) = \Braket{\hat{b}_i^\dagger \hat{b}_{i+r}}. 
\end{gather}
and evaluate them using the procedure outlined above for the DW correlation functions (see the supplemental material \cite{SM} for more details).
The superfluid phase in one dimension exhibits only quasi-long range order and is characterized by a power law decay of the correlation function with distance. On the other hand, $C_{SF}$ decays exponentially in the insulating phases. }
If the ratio between kinetic energy and interaction energy is sufficiently different between the two bands, we may obtain an interesting scenario where the strongly interacting band supports insulating behavior, whereas the weakly interacting band favors a SF state. Specifically, we are interested in the case where strong nearest-neighbor interactions ($V>U/2$) produce the DW order of the insulating state.
We have studied this case using the DMRG method to calculate the ground state\footnote{In our DMRG calculation, the maximum bond dimension is set to $800$. As a criterion for convergence, the discrepancy between subsequent iterations shall be below $10^{-8}$ for energy and $10^{-5}$ for entropy. 
}. 
\textcolor{black}{
Although DMRG works most efficiently under open boundary conditions, we have used periodic boundary conditions in the following calculations, since in the DW phase open boundary conditions require an artificial potential to break the degeneracy.
}
 The phases are identified by evaluating the DW order parameter and the SF correlation function for the two bands separately, as well as for the entire system as a whole.
 For a broad range of interband interaction parameters $U_{ab}$ and $V_{ab}$,  our results are illustrated and summarized in Fig.~\ref{fig:phasediagram} for a fixed choice for intraband parameters. The color bar shows the value of the density wave order parameter.
The red or blue circles indicate the qualitative behavior of the SF correlations; blue circles correspond to exponential decay (no SF), whereas red circles correspond to algebraic decay (SF). 
\textcolor{black}{
If any of the interband interactions $U_{ab}$ or $V_{ab}$ becomes large, the interactions separate the system into two different phases.
In the diagram of Fig.~\ref{fig:phasediagram}, the phase separation is distinguished by the particle number profiles and assigned to the white region.
}

Concentrating on the interesting non-phase separated regime  (colored regions in Fig.~\ref{fig:phasediagram} ), we first analyze the two bands separately. We find that the strongly interacting band (band A) exhibits DW order and no SF order (left panel). The behavior of band B (middle panel) is more diverse: Unless $U_{ab}$ or $V_{ab}$ get very large, this band exhibits SF order, as indicated by the algebraic decay of SF correlators. 
Since the repulsive $U_{ab}$ ($V_{ab}$) has a tendency of correlating (anti-correlating) the density of band B with the DW-ordered density of band A, both types of interband interactions have a tendency of producing DW order in band B, unless both $U_{ab}$ and $V_{ab}$ get so strong that phase separation is reached. The simultaneous presence of SF and DW order establishes supersolid (SS) behavior of band B. However, if one of the interband interactions becomes very strong,  band B becomes insulating, with exponentially decaying SF correlators  and finite DW order. As already explained in Fig.~\ref{fig:5PossibleState}, the two interband interactions $U_{ab}$ and $V_{ab}$ lead to different combination of two DW patterns. Hence, in an intermediate regime,  $U_{ab} \approx 2V_{ab}$, the two opposite effects cancel each other, and band B remains with uniform density. 
The different effect of the two interaction types is best seen when considering both bands together (right panel): We note that the regime in which $U_{ab}$ dominates and densities of band A and B are anti-correlated (denoted SS1 and DW1 in the middle panel), the overall density is homogeneous. On the other hand, the regime with correlated densities (SS2 and DW2), exhibit density modulations in the entire system. We emphasize that, no matter if the overall density is homogeneous or modulated, the simultaneous occurrence of SF order and spatial symmetry breaking establishes some sort of SS behavior.

\begin{figure}
    \centering
    \includegraphics[scale = 0.5]{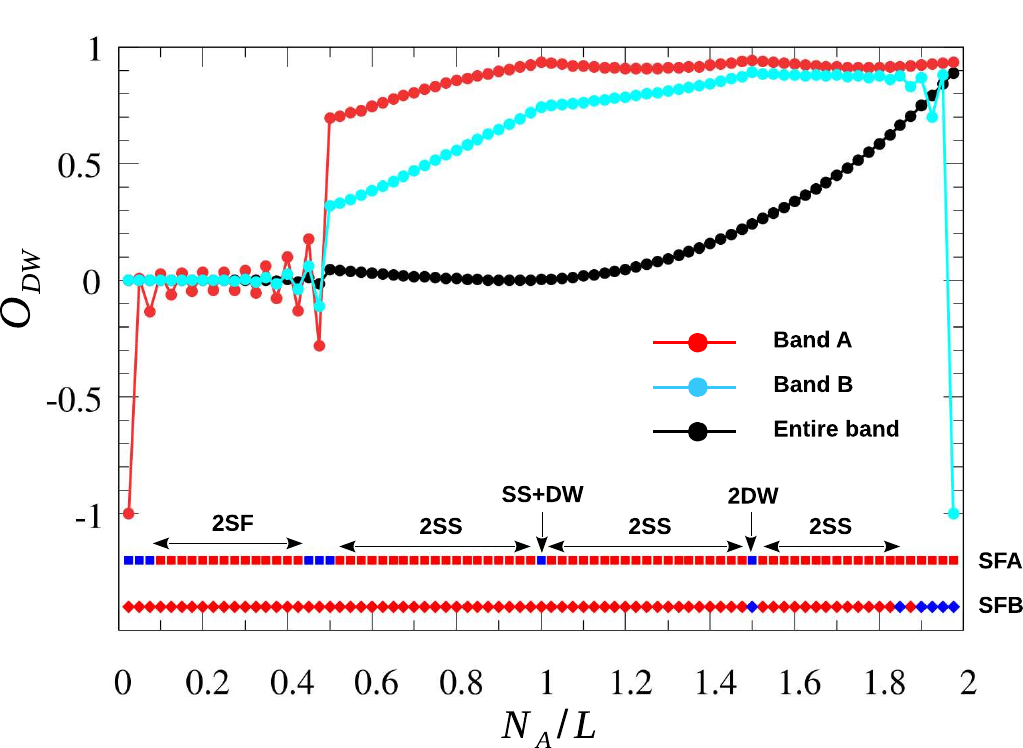}
    \caption
    {
    The dependence of the density wave order parameter and the superfluid correlation function on the particle population in the SS1 regime $U_{ab} = 3.2,\ V_{ab} = 0$.
    The circle lines show the density wave order parameter in the lower (red), the higher (light-blue), and the entire (black) band. 
    The red and blue rectangles (circles) show the algebraic and exponential decay of the superfluid correlation function in the lower (or the upper) band, respectively.}
    \label{fig:population_SS1}
\end{figure}

\textit{Polarization dependence of the phase.} While synthetic systems like cold atomic setups may allow to freely choose the band polarization, in other systems the occupation of the bands is governed by the temperature and the energy gap between the bands. This includes the relevant case of excitons in synthetic lattices, where finite temperature, a tunable band gap, and strong interactions can result in the sizable occupation of excited bands \cite{Lagoin2022}. Due to the more delocalized nature of excited Wannier states, the excited band has a significantly larger hopping parameter. This provides strong motivation for studying the interplay of an interaction-dominated DW band and a tunneling-dominated SF band beyond the unpolarized case. Therefore, in the remainder of this Letter we ask the question whether the recipe of generating SS order through the interplay of such bands is generic, or whether it is limited to the unpolarized case.

To answer this question, we concentrate on a fixed point in the parameter space (chosen either from the SS1 or SS2 regime determined in Fig.~\ref{fig:phasediagram}), and analyze the ground states in the different polarization sectors from $(N_a=0, N_b=2L)$ to $(N_a=2L,N_b=0)$ a. The results are exemplified in Fig.~\ref{fig:population_SS1} for a data point from the SS1 regime.
In the figure, we plot the DW order parameter as a function of the population $N_a$. In order to establish DW order, band A needs to be at least half filled ($N_a \geq L/2$). Under this condition, also band B is found to establish DW order. 
Interestingly, for larger $N_a$, the DW behavior is seen both separately in each band as well as in the total density of both bands together whose behaviour originates from the anticorrelation of the two in general unequal DWs which are shifted w.r.t each other by one site. 

In the lower part of the plot in Fig.~\ref{fig:population_SS1}, we also indicate the decay behavior of the SF correlators in the two bands by the red and blue dots. Algebraic decay (red dots) is found to be the prevalent behavior in both bands unless their population drops to very small values. However, in the vicinity of special fillings (specifically $N_a/L=1/2, 1$, and $3/2$) the insulating DW configuration is stabilized. The enhanced DW order is also seen in small cusps of $O_{DW}^A$. At these fillings, the DW order prevents the system from establishing SF order in band A. For $N_a/L=3/2$, the suppression of SF order also affects band B.

We mention that a similar behavior has been observed in the SS2 regime for $N_a>L/2$. As expected for this regime, the DW order of the two bands interfere constructively. 
For small $N_a<L/2$, the appearance of phase separation in the SS2 regime constitutes a major difference from the SS1 case.
\begin{figure}[!tbp]
    \centering
    \includegraphics[width =\columnwidth]{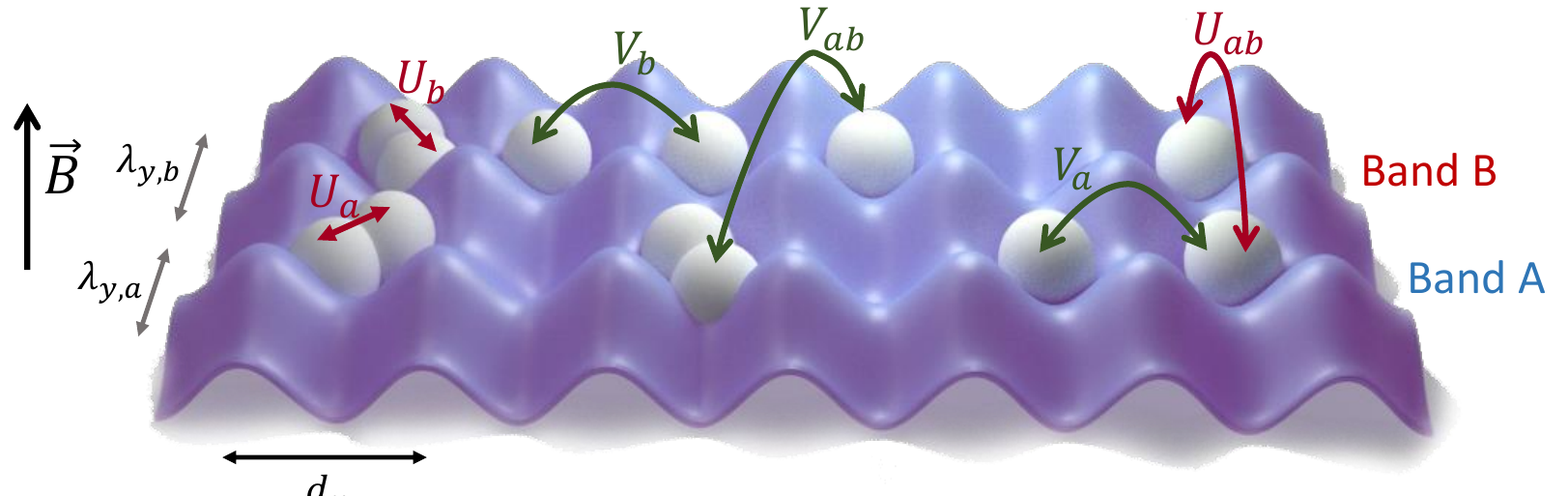}
    \caption{Schematic representation for the implementation of the extended Bose-Hubbard model. Bands A/B are associated to different layers of the optical lattice. $U_{a/b}$ terms correspond to on-site atomic interactions, and extended forces $V_{a/b},\, U_{ab},\,V_{ab}$ are induced by dipolar interactions between sites separated along the horizontal, vertical and diagonal directions, respectively.}
    \label{fig:bilayer}
\end{figure}

\textit{Implementation with cold atoms.} {\color{black} A highly flexible candidate for the experimental realization of the Bose-Hubbard model in Eq.~\eqref{eq:hamiltonian} are cold atoms in a double-ladder optical lattice. One can map the occupation of each band to the presence of a bosonic atom on each of the ladders of a double-ladder optical lattice. Hopping $J_{a(b)}$ is thus associated to atomic tunneling along the horizontal(vertical) axis and $U_{a(b)}$ corresponds to contact interactions, which can be independently tuned by choosing a different trap depth for each ladder, as it occurs in bipartite lattices created by the combination of two laser frequencies along the vertical direction~\cite{lohseThouless2016, modugnoMaximally2012}. This also allows one to induce an energy offset $\Delta$ between the two ladders. Terms $U_{ab}$, $V_{a(b)}$ and $V_{ab}$ describe repulsive interactions along the nearest-neighbor vertical, horizontal and diagonal directions, respectively (see Fig. \ref{fig:bilayer}). Engineering such non-local terms can benefit from recent experimental achievement with magnetic atoms such as erbium~\cite{chomazDipolar2022,Su2023}, where a lattice spacing of order $250$ nm translates into strong dipolar interactions that are comparable to the tunneling frequency~\cite{SM} (see also the references \cite{guardado-sanchezQuench2021, cornishQuantum2024, baierExtended2016, patscheiderDetermination2022} therein). Dipolar forces highly depend on the atomic separation, and the relative polar angle of the dipole. This allows one to highly engineer the density-density terms by appropriately choosing the lattice spacing and atomic quantization axis, while longer-range interactions decay polynomially. 
Thanks to current advances with accordion lattices~\cite{Su2023}, an atomic gas microscope can resolve the position of each individual atom, confirming the polarization of the system and measuring the DW correlation functions in Eqs.~(\ref{eq:DWcorrelators1},\ref{eq:DWcorrelators2}) through fluorescence measurements. Time-of-flight measurements can also discriminate between superfluid and insulating phases~\cite{Greiner2002}.}

\textit{Summary.} Our findings establish the existence of SF, SS, DW insulating phases, as well as phase separation in two-band extended Bose-Hubbard models. These different phases can be selected by tuning the interband interaction parameters, as we have explicitly shown for an unpolarized system at filling one, but also at fixed system parameters by varying the band polarization. Our results provide a recipe for achieving supersolidity by combining a band with SF properties and a band with DW properties, e.g. by thermally exciting a second band. Interestingly, no matter which band provides the majority of particles, the combination of the two unequal bands can result in SS order.

We emphasize that our study of a two-band extended Bose-Hubbard chain does not exhaust the rich opportunities which arise from the combination of orbital degrees of freedom and long-range interactions. For instance, the combination of two superfluid bands may also lead to twisted multi-orbital superfluidity, similar to the one studied in Refs.~\onlinecite{Soltan-Panahi2012,Juergensen2015}, or topological superfluidity \cite{Lv2016,Wang2021}. Another interesting possibility would be to look at effective higher-body interactions if interactions beyond the density-density produce a virtual occupation of one band \cite{Mazets2008,Will2010}.

\textit{Acknowledgements}
 T.~G. acknowledges funding by Gipuzkoa Provincial Council (QUAN-000021-01), by the Department of Education of the Basque Government through the IKUR strategy and through the project PIBA\_2023\_1\_0021 (TENINT), by the Agencia Estatal de Investigaci{\' o}n (AEI) through Proyectos de Generaci{\' o}n de Conocimiento PID2022-142308NA-I00 (EXQUSMI), by the BBVA Foundation (Beca Leonardo a Investigadores en F{\' i}sica 2023). The BBVA Foundation is not responsible for the opinions, comments and contents included in the project and/or the results derived therefrom, which are the total and absolute responsibility of the authors.
  U.B. acknowledges the project
that gave rise to these results, received the support of a fellowship (funded from the European Union’s Horizon 2020 research and innovation programme under the Marie Sklodowska-Curie grant agreement No 847648) from “la Caixa” Foundation (ID 100010434). The fellowship code is “LCF/BQ/PR23/11980043”. RWC   acknowledges support from the Polish National Science Centre (NCN) under the Maestro Grant No. DEC- 2019/34/A/ST2/00081.
ICFO group acknowledges support from: ERC AdG NOQIA; MICIN/AEI (PGC2018-0910.13039/501100011033,  CEX2019-000910-S/10.13039/501100011033, Plan National FIDEUA PID2019-106901GB-I00, FPI; MICIIN with funding from European Union NextGenerationEU (PRTR-C17.I1): QUANTERA MAQS PCI2019-111828-2); MCIN/AEI/10.13039/501100011033 and by the “European Union NextGeneration EU/PRTR"  QUANTERA DYNAMITE PCI2022-132919 (QuantERA II Programme co-funded by European Union’s Horizon 2020 programme under Grant Agreement No 101017733), Proyectos de I+D+I “Retos Colaboración” QUSPIN RTC2019-007196-7); Fundació Cellex; Fundació Mir-Puig; Generalitat de Catalunya (European Social Fund FEDER and CERCA program, AGAUR Grant No. 2021 SGR 01452, QuantumCAT \ U16-011424, co-funded by ERDF Operational Program of Catalonia 2014-2020); Barcelona Supercomputing Center MareNostrum (FI-2023-1-0013); EU Quantum Flagship (PASQuanS2.1, 101113690); EU Horizon 2020 FET-OPEN OPTOlogic (Grant No 899794); EU Horizon Europe Program (Grant Agreement 101080086 — NeQST), National Science Centre, Poland (Symfonia Grant No. 2016/20/W/ST4/00314); ICFO Internal “QuantumGaudi” project; European Union’s Horizon 2020 research and innovation program under the Marie-Skłodowska-Curie grant agreement No 101029393 (STREDCH) and No 847648  (“La Caixa” Junior Leaders fellowships ID100010434: LCF/BQ/PI19/11690013, LCF/BQ/PI20/11760031,  LCF/BQ/PR20/11770012, LCF/BQ/PR21/11840013). Views and opinions expressed are, however, those of the author(s) only and do not necessarily reflect those of the European Union, European Commission, European Climate, Infrastructure and Environment Executive Agency (CINEA), nor any other granting authority.  Neither the European Union nor any granting authority can be held responsible for them. 

\bibliography{bibliography.bib}


\end{document}